\documentclass{moriond}

\usepackage{amsmath,amssymb}

\def\be{\begin{equation}}
\def\ee{\end{equation}}
\def\bea{\begin{eqnarray}}
\def\eea{\end{eqnarray}}

\newcommand{\Photo}{}

\begin{document}

\rightline{DO-TH 16/10, QFET-2016-08}

\vspace*{4cm}
\title{OPPORTUNITIES WITH (SEMI)LEPTONIC RARE CHARM DECAYS}

\author{ STEFAN DE BOER }

\address{Fakult\"at f\"ur Physik,\\
TU Dortmund, Otto-Hahn-Str.4, D-44221 Dortmund, Germany}

\maketitle\abstracts{
We study (semi)leptonic rare charm decays and its opportunities in searches for physics beyond the Standard Model (BSM).
In particular, we analyse the impact of potential BSM physics in $c\to ull'$ transitions, notably branching ratios, angular observables, asymmetries and Lepton Flavour Violating (LFV) decays.
Testable effects are worked out model-independently and within Leptoquark models supplemented with flavour patterns to link $K$/$B$ decays.}

\section{Introduction}

Flavour Changing Neutral Current (FCNC) induced (semi)leptonic charm decays are rare in the SM due to an effective Glashow-Iliopoulos-Maiani (GIM) mechanism~\cite{Glashow:1970gm} and additionally loop-suppressed, thus sensitive to BSM physics.
On the experimental side smaller upper limits on the branching ratios have been set, notably on the non-resonant mode $\mathcal B^\text{nr}(D^+\to\pi^+\mu^+\mu^-)<7.3\times10^{-8}$~\cite{Aaij:2013sua} and recently $\mathcal B(D^0\to e^\pm\mu^\mp)<1.3\times10^{-8}$~\cite{Aaij:2015qmj}.
On the theoretical side the convergence of calculations by means of $\alpha_s$ and $\Lambda_\text{QCD}/m_c$ is questionable.
In particular, rare charm decays are unique up-type quark FCNC transitions complementary to $K$/$B$ physics.

We present opportunities with (semi)leptonic rare charm decays in searches for BSM physics, that is complementary observables model-independently and within Leptoquark models supplemented with flavour patterns to link $K$/$B$ decays based on reference~\cite{deBoer:2015boa}.
In the next section we present the observables one by one, emphasising its potential and downside, each.

\section{Opportunities with (semi)leptonic rare charm decays}

\begin{figure}[ht]
\begin{minipage}{0.5\linewidth}
\centerline{\includegraphics[width=0.8\linewidth]{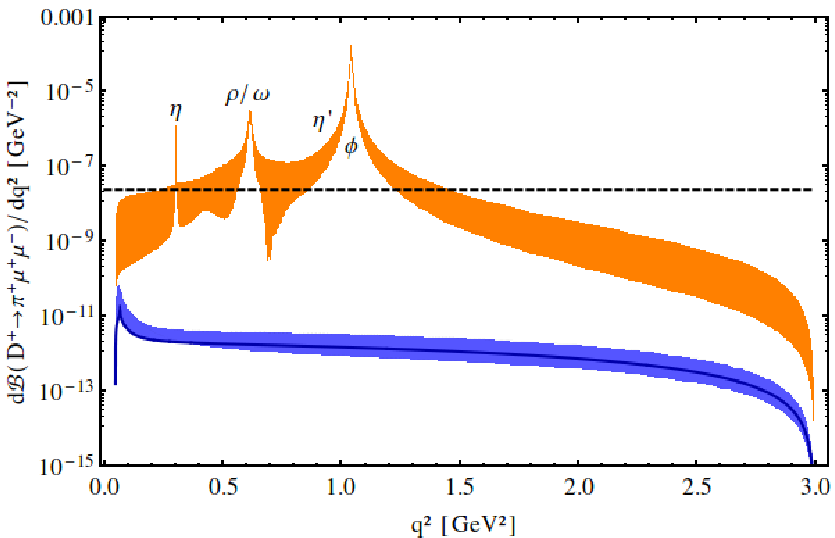}}
\end{minipage}
\hfill
\begin{minipage}{0.5\linewidth}
\centerline{\includegraphics[width=0.8\linewidth]{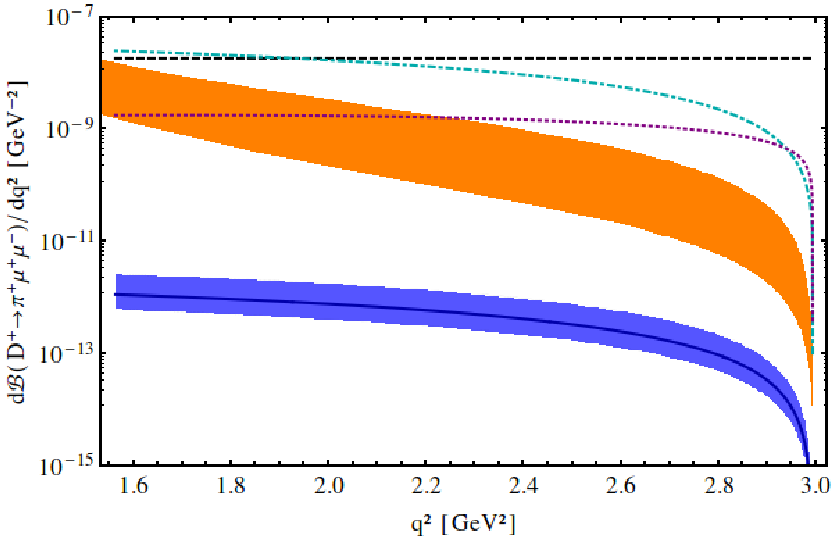}}
\end{minipage}
\caption[]{The $q^2$ distribution of the branching ratio of $D^+\to\pi^+\mu^+\mu^-$.
The dashed black line is the binned 90\% CL experimental upper limit.
The orange band represents the resonant modes modelled via a Breit-Wigner shape to fit the data and varying the relative strong phases, the solid blue curve is the non-resonant SM (next-to)next-to-leading order QCD prediction within an operator product expansion at $\mu_c=m_c$ and the lighter blue band its $\mu_c$-uncertainty.
The right plot shows $q^2\ge(1.25\,\text{GeV})^2$ and two additional (dot-dashed cyan and dotted purple) curves for potential BSM Wilson coefficients.
Figures taken from reference~\cite{deBoer:2015boa}.}
\label{fig:dBQDppiR2mu}
\end{figure}

First, we have a look at the dilepton mass ($q^2$) decay distribution of $D^+\to\pi^+\mu^+\mu^-$ in Figure~\ref{fig:dBQDppiR2mu} (left plot).
We see that the contributions from resonant modes are larger than the non-resonant SM branching ratio.
At high $q^2$, above the $\phi$ resonance, the experimental upper limit is above the resonant branching ratio, thus opening a window to make BSM effects visible.
A closer look at Figure~\ref{fig:dBQDppiR2mu} (right plot) shows two distributions within a model-independent BSM approach consistent with the experimental limit on $\mathcal B(D^0\to\mu^+\mu^-)$ above the resonant branching ratio.
On the downside BSM effects have to be large to be observed in the branching ratio, guiding us to approximative SM null-test observables.

Second, semileptonic angular observables are defined via
\begin{equation}
\frac1\Gamma\frac{\mathrm d\Gamma}{\mathrm d\cos\theta}=\frac34(1-F_H)(1-\cos^2\theta)+A_{FB}\cos\theta+\frac12F_H\,,
\end{equation}
where $\theta$ is the angle between the $l^-$ and the $D^+$ in the dilepton center-of-mass frame.
Within the model-independent BSM approach at high $q^2$ the forward-backward asymmetry is constrained to be
\begin{equation}
|A_{\rm FB}(D^+\to\pi^+\mu^+\mu^-)|\lesssim0.6
\end{equation}
and the flat term
\begin{equation}
F_H(D^+\to\pi^+\mu^+\mu^-)\lesssim1.5\,.
\end{equation}
The angular observables vanish in the SM.
Additionally, the semileptonic decays $D^+\to\pi^+e^\pm\mu^\mp$ and $D^+\to\pi^+\nu\bar\nu$ are SM null-tests.
Within the model-independent BSM approach the LFV and dineutrino branching ratios may be close to the experimental limit $\mathcal B(D^+\to\pi^+e^\pm\mu^\mp)\lesssim3\cdot10^{-6}\;\text{@CL=90\%}$~\cite{Lees:2011hb} and $\mathcal B(D^+\to\pi^+\nu\bar\nu)\sim10^{-5}$, where we suppose the dineutrino branching ratio to be observable at BESIII~\cite{Asner:2008nq}.
Thus, any non-zero measurement of any of the branching ratios or the angular observables is due to BSM effects.
On the downside no correlations to current observables measured experimentally exist, guiding us towards a model-dependent BSM approach.

As a BSM scenario, we take leptoquark models that induce $c\to u$ transitions, where its quantum numbers are given in Table~\ref{tab:LQ_qn}.
Leptoquark models are of recent interest as they may generate Lepton Non-Universality (LNU) in $R(K)$ and $R(D^*)$~\cite{Alonso:2015sja,Bauer:2015knc,Fajfer:2015ycq} and may induce the 750 GeV diphoton decay~\cite{Bauer:2015boy,Murphy:2015kag}.
We order them into two classes, where leptoquarks coupling to quark singlets are labelled case (1) and leptoquarks coupling to quark doublets are labelled case (2).
The couplings and masses of the leptoquarks are constrained by collider experiments and, e.g. $\mu\to e\gamma$ and $\mu-e$ conversion in nuclei and observables in $K$ physics for case (2).~\cite{deBoer:2015boa}
Additionally, we correlate the couplings via flavour patterns inspired by Frogatt-Nielsen $U(1)$ (quarks, rows) and $A_4$ (leptons, columns) symmetries, e.g.~\cite{Varzielas:2015iva}
\begin{equation}
\lambda_\mathrm{i,ii,iii}\sim\left(
\begin{array}{ccc}
\rho_d\kappa  &  \rho_d  &  \rho_d  \\
\rho\kappa    &  \rho    &  \rho  \\
\kappa        &  1       &  1  \\
\end{array}
\right)\,,\quad\left(
\begin{array}{ccc}
0  &  *  &  0  \\
0  &  *  &  0  \\
0  &  *  &  0  \\
\end{array}
\right)\,,\quad\left(
\begin{array}{ccc}
*  &  0  &  0  \\
0  &  *  &  0  \\
0  &  *  &  0  \\
\end{array}
\right)\,,
\end{equation}
where $\lambda_\mathrm{i}$ is a hierarchical flavour pattern, $\lambda_\mathrm{ii}$ is a single lepton pattern and $\lambda_\mathrm{iii}$ is a first two generation diagonal pattern.
\begin{table}[ht]
\caption[]{Scalar and vector leptoquarks (LQ) and their quantum numbers $(SU(3)_C,SU(2)_L,Y)$.}
\label{tab:LQ_qn}
\vspace{0.4cm}
\begin{center}
\begin{tabular}{|c|c|}
\hline
$S$calar LQ      &  $V$ector LQ \\
\hline
$S_1(3,1,-1/3)$  &  $\tilde V_1(3,1,-5/3)$  \\
$S_2(3,2,-7/6)$  &  $V_2(3,2,-5/6)$  \\
$S_3(3,3,-1/3)$  &  $\tilde V_2(3,2,1/6)$  \\
                 &  $V_3(3,3,-2/3)$  \\
\hline
\end{tabular}
\end{center}
\end{table}

Third, leptoquark models correlated via flavour patterns and consistent with current constraints yield the branching ratios of semileptonic and leptonic modes given in Table~\ref{tab:LQ_B}, where the branching ratios of $c\to ue^+e^-$ modes are SM-like, thus LNU in charm decays may be generated.
Note that $\mathcal B(D^0\to\mu^\pm e^\mp)$ as measured recently tests the leptoquark models and flavour patterns.
On the downside correlations to $K$/$B$ decays are not measurable, guiding us to a resonance catalysed observable~\cite{Fajfer:2012nr}.
\begin{table}[ht]
\caption[]{Branching ratios on the full $q^2$-bin (high $q^2$-bin) for two classes of leptoquark models supplemented with flavour patterns and its experimental sensitivity~\cite{Aaij:2013cza}.
The complete table can be found in~\cite{deBoer:2015boa}.}
\label{tab:LQ_B}
\vspace{0.4cm}
\begin{center}
\begin{tabular}{|c|c|c|}
\hline
         &  $\mathcal B(D^+\to\pi^+\mu^+\mu^-)$        &  $\mathcal B (D^0\to\mu^+\mu^-)$  \\
\hline
(ii.1)   &  $\lesssim7\cdot10^{-8}$ ($2\cdot10^{-8}$)  &  $\lesssim3\cdot10^{-9}$  \\
(iii.1)  &  SM-like                                    &  SM-like  \\
exp.     &  $<7.3\cdot10^{-8}$ ($2.6\cdot10^{-8}$)     &  $<6.2\cdot10^{-9}$  \\
\hline
\end{tabular}

\vspace{0.4cm}
\begin{tabular}{|c|c|c|c|}
\hline
         &  $\mathcal B(D^+\to\pi^+e^\pm\mu^\mp)$  &  $\mathcal B(D^0\to\mu^\pm e^\mp)$  &  $\mathcal B(D^+\to\pi^+\nu\bar\nu)$  \\
\hline
(ii.1)   &  $0$                                    &  $0$                                &  $\lesssim8\cdot10^{-8}$  \\
(iii.1)  &  $\lesssim2\cdot10^{-6}$                &  $\lesssim4\cdot10^{-8}$            &  $\lesssim 2 \cdot 10^{-6}$  \\
exp.     &  $\lesssim3\cdot10^{-6}$                &  $<1.3\cdot10^{-8}$                 &  $\sim10^{-5}$  \\
\hline
\end{tabular}
\end{center}
\end{table}

\begin{figure}[ht]
\begin{minipage}{0.5\linewidth}
\centerline{\includegraphics[width=0.8\linewidth]{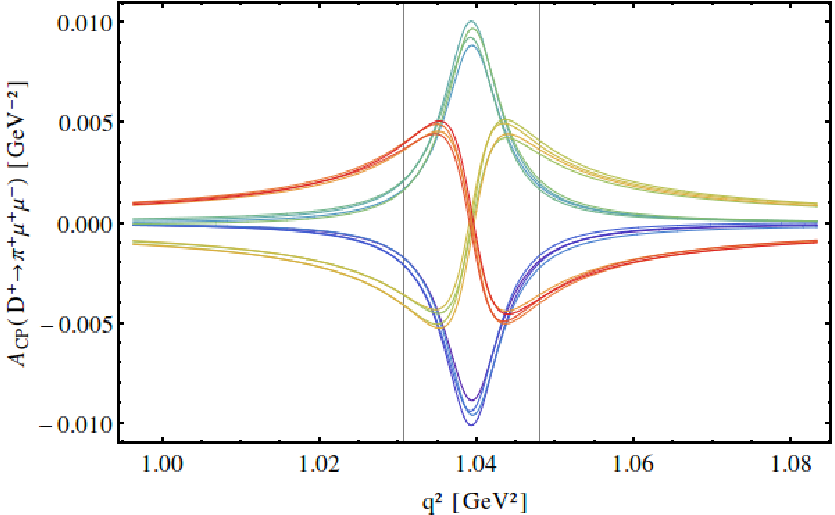}}
\end{minipage}
\hfill
\begin{minipage}{0.5\linewidth}
\centerline{\includegraphics[width=0.8\linewidth]{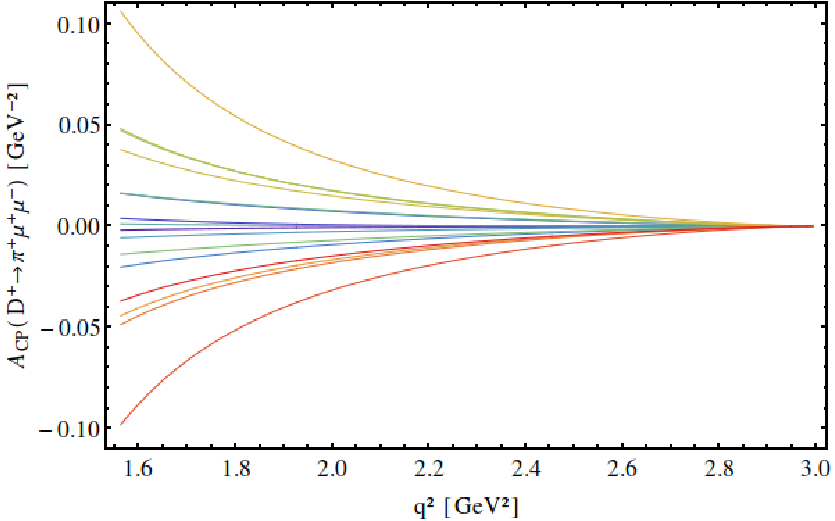}}
\end{minipage}
\caption[]{The $CP$ asymmetry normalized to the shown bins for case (ii.2) around the $\phi$ resonance (left plot) and at high $q^2$ (right plot).
From yellow (upper curves above $\phi$) to red (lower curves above $\phi$) each bunch represents the resonant phase $\delta_\phi=\pi/2,\pi,0,3/2\pi$.
The vertical lines are $(m_\phi\pm\Gamma_\phi)^2$.
Figures taken from reference~\cite{deBoer:2015boa}, where additional plots for case (ii.1) can be found.}
\label{fig:ACPphirhoDppi2mu}
\end{figure}

Fourth, the $CP$ asymmetry is defined as
\begin{align}
A_{CP}(q^2)=\frac{\mathrm d\Gamma/\mathrm dq^2-\mathrm d\bar\Gamma/\mathrm dq^2}{\int_{q^2_\text{min}}^{q^2_\text{max}}\mathrm dq^2(\mathrm d\Gamma/\mathrm dq^2+\mathrm d\bar\Gamma/\mathrm dq^2)}\,,
\end{align}
where $\mathrm d\bar\Gamma/\mathrm dq^2$ is the rate distribution of the $CP$-conjugated mode, $D^-\to\pi^-l^+l^-$.
The $CP$ asymmetry, negligible in the SM, for leptoquark models supplemented with flavour patterns and consistent with current constraints is shown in Fig.~\ref{fig:ACPphirhoDppi2mu}.
Around the $\phi$ resonance (left plot) the $CP$ asymmetry is sensitive to BSM physics generating the operator $Q_9=(\bar u_L\gamma_\mu c_L)(\bar l\gamma^\mu l)$ independent of the resonant phases.
At high $q^2$ (right plot) a small BSM induced Wilson coefficient $C_9$, e.g. as linked to $K$/$B$ physics, may induce larger $CP$ asymmetries.

\section{Conclusion}

We have presented BSM opportunities with (semi)leptonic rare charm decays.
Notably, $\mathcal B(D^+\to\pi^+\mu^+\mu^-)$ above the $\phi$-resonance, angular observables, $CP$ asymmetries, LFV and dineutrino decays are complementary to search for potential BSM physics.
Additionally, leptoquark models link charm and $K$/$B$ physics, e.g. LNU, thus flavour models are testable.

\section*{Acknowledgements}

I would like to thank the organizers for the invitation and the wonderful conference.
I am grateful to Gudrun Hiller for a fruitful collaboration and Kamila Kowalska for reading the manuscript.
The work reported has been supported in part by the DFG Research Unit FOR 1873 ``Quark Flavour Physics and Effective Field Theories''.


\end{document}